\documentclass[
 twocolumn,
 nofootinbib,
 amsmath,amssymb,
 aps,
 prd,
 superscriptaddress,
]{revtex4-2}

\usepackage{xcolor}
\usepackage[colorlinks=true, linkcolor=red, citecolor=blue, urlcolor=blue]{hyperref}
\usepackage{graphicx}
\usepackage{bm}
\usepackage{dcolumn}
\usepackage{booktabs}
\usepackage{multirow}
\usepackage{mathtools}
\usepackage{ulem}

\newcommand{\MBH}{M_{\rm PBH}}
\newcommand{\MH}{M_{\rm H}}

\newcommand{\dd}{\mathrm{d}}
\newcommand{\Pzeta}{\mathcal{P}_{\zeta}}
\newcommand{\Pdelta}{\mathcal{P}_{\delta}}
\newcommand{\deltath}{\delta_{\rm th}}

\newcommand{\deltaHth}{\delta_{{\rm H},{\rm th}}}

\newcommand{\Adss}{A_{\rm DSS}}
\newcommand{\Pln}{P_{\ln}}
\newcommand{\eps}{\epsilon}

\newcommand{\maybeincludegraphics}[2][]{%
  \IfFileExists{#2}{%
    \includegraphics[#1]{#2}%
  }{%
    \fbox{%
      \begin{minipage}[c][0.23\textheight][c]{0.92\linewidth}
      \centering
      Missing figure file:\\[0.4em]
      \texttt{\detokenize{#2}}\\[0.4em]
      Upload this file to Overleaf or replace the filename.
      \end{minipage}}
  }%
}

\begin{document}

\preprint{RUP-26-XX}

\title{Discrete self-similarity imprints on primordial-black-hole mass functions}

\author{Luis E. Padilla}
\affiliation{Department of Physics, Rikkyo University, Tokyo, Japan}

\author{Tomohiro Harada}
\affiliation{Department of Physics, Rikkyo University, Tokyo, Japan}

\author{Hayami Iizuka}
\affiliation{Department of Physics, Rikkyo University, Tokyo, Japan}
\date{\today}

% ============================================================
\begin{abstract}
We study how discrete self-similarity (DSS) in the critical behavior of scalar-field collapse is imprinted on primordial-black-hole (PBH) mass functions. Using the DSS-modulated critical scaling law found in cosmological simulations, we propagate the near-threshold mass map into normalized PBH mass functions during a {kination} era. We compare Gaussian window function, $k$-space top-hat window function, and real-space top-hat window function, together with a {real-space top-hat window function multiplied by a kination transfer function}. We find that critical behavior in gravitational collapse produces an irreducible minimum width even for an infinitesimally narrow primordial spectrum. DSS then modulates this critical-scaling profile, generating approximately log-periodic features in mass. At fixed horizon mass, successive equal-phase points of the DSS-modulated critical mass map satisfy
$\Delta\ln \MBH=\gamma\Pln$. Consequently, in the narrow-spectrum limit, the corresponding structures in the final mass function are expected to satisfy $\Delta\ln m\simeq\gamma\Pln$, or $m_{n+1}/m_n\simeq5.6$, for the fiducial Choptuik DSS parameters {$\gamma$ and $\Pln$}. For broad primordial spectra, the convolution over horizon masses dephases the DSS pattern and progressively washes out the critical substructure. {We normalize the mass functions so that we may primarily study} the profile shape and the survival of DSS substructure, rather than the absolute PBH abundance. These results provide a bridge between cosmological DSS collapse simulations and PBH {phenomenology including} population observables.
\end{abstract}


\maketitle

% ============================================================
\section{Introduction}
\label{sec:introduction}

Primordial black holes (PBHs) provide a unique probe of the early Universe, since their formation depends on the rare, nonlinear tail of primordial perturbations and on strong-field gravitational dynamics. In many scenarios PBHs form when large-amplitude primordial curvature perturbations re-enter the horizon and collapse against pressure gradients and cosmological expansion \cite{Zeldovich:1967lct,Hawking:1971ei,Carr:1974nx,Carr:2020gox,Green:2020jor,Sasaki:2018dmp,Byrnes:2025tji}. Their abundance and mass distribution are therefore sensitive not only to the primordial curvature power spectrum, but also to the nonlinear mapping between the amplitude of a primordial fluctuation and the final black-hole mass.

A central ingredient of this mapping is critical behavior. Close to the threshold of black-hole formation, the final PBH mass follows the {power-law} critical-scaling relation
\begin{equation}
\MBH=\mathcal{K}\MH(\delta-\deltath)^\gamma,
\label{eq:intro_standard_scaling}
\end{equation}
where $\delta$ is a perturbation-amplitude variable, $\deltath$ is the corresponding collapse threshold, $\MH$ is the horizon mass of the collapsing scale, $\mathcal{K}$ is a normalization constant, and $\gamma$ is the critical exponent {~\cite{Evans:1994pj,
Niemeyer:1997mt,Niemeyer:1999ak,Musco:2012au}.
This relation is well understood by the renormalisation group flow of a 
continuously self-similar (CSS) solution at the threshold, which is called a critical solution, acting as an intermediate attractor~\cite{Koike:1995jm,Maison:1995cc,Koike:1999eg}.} 
Equation~\eqref{eq:intro_standard_scaling} has an immediate consequence for PBH phenomenology: even if the primordial power spectrum were sharply localized around a single scale, PBHs would not form with exactly one mass. At fixed horizon mass, a continuous range of supercritical amplitudes above threshold is mapped into a continuous range of PBH masses. The power-law critical-scaling relation therefore provides an irreducible source of width in the PBH mass function.

For a massless scalar field, the critical solution in asymptotically flat spacetime is discretely self-similar (DSS), {instead of} {CSS} \cite{Choptuik:1992jv}. The corresponding mass scaling is therefore not a pure power law: it contains a log-periodic residual whose period is related to the echoing period of the DSS critical solution \cite{Hod:1996az,Gundlach:1997gc}. In PBH formation, this implies a log-periodic modulation of the critical mass map as a function of \(\ln M_{\rm PBH}\), which can generate approximately log-periodic substructure in the final mass function as a function of \(\ln m\).

The massless scalar-field model is particularly interesting for two independent
reasons. First, it is the original and cleanest system in which critical
collapse and DSS were discovered. It therefore provides the natural
theoretical laboratory in which { we} ask how DSS modifies PBH observables.
Second, the same system is physically relevant in cosmology. A
kination phase corresponds to a period in which the potential energy
of a scalar field is negligible and the energy density is dominated by the
kinetic term. In this regime, as long as the scalar-field gradient is timelike, the stress-energy
tensor is equivalent to that of a perfect fluid with equation of state $w=1$.
Kination phases arise naturally in models of quintessential inflation,
curvaton dynamics, and post-inflationary reheating
\cite{Peebles:1998qn,Gouttenoire:2021jhk}. More generally,
scalar-field-dominated post-inflationary epochs can support several
distinct channels of PBH formation, whose collapse dynamics and mass
mapping depend on the scalar-field evolution
\cite{Padilla:2025bkv,Milligan:2025zbu,Padilla:2024iyr,Padilla:2021zgm}. Thus, the massless scalar-field
model is both the canonical setting for DSS and a physically motivated
description of scalar-field-dominated early-Universe phases beyond radiation
domination.

From this point of view, kination is not only a {well-}motivated cosmological scenario, but also a useful case study for a broader question: can the critical behavior of gravitational collapse leave an imprint on PBH population observables? In the massless scalar-field case, the answer is particularly sharp because the critical solution is discretely self-similar, leading to a log-periodic structure in the mass map. Other early-Universe matter sectors or equations of state may exhibit different critical behavior and therefore different residual patterns. The DSS signal studied here should therefore be viewed as one example of a more general possibility: PBH observables, such as the PBH mass function, may retain information about the critical behavior of gravitational collapse, not only about the primordial curvature spectrum.

Recently, fully general-relativistic simulations of cosmological massless scalar-field collapse showed that DSS survives in an expanding FLRW background \cite{Padilla:2026dss}. In that work, the critical PBH mass-scaling relation was found to contain clear log-periodic modulations, with effective values close to the standard asymptotically flat Choptuik values. This raises the central question addressed in the present paper: if the local near-threshold mass map is DSS-modulated, does this {DSS modulation} survive after constructing the full PBH mass function?

The answer is not obvious. The PBH mass function is not determined by the critical scaling relation alone; it is obtained after integrating over perturbation amplitudes, smoothing scales, horizon masses, window functions, and the shape of the primordial power spectrum. A broad primordial spectrum can mix contributions from many horizon masses and many phases of the DSS oscillation. As a result, the DSS signal may be visible, distorted, or erased depending on the width of the primordial spectrum and on the {choice of window function}. In this work we use the DSS-modulated mass law motivated by Ref.~\cite{Padilla:2026dss} to study this propagation explicitly.

Previous work provides two important pieces of context for this question. The first is the role of critical behavior in PBH mass-function calculations. Gow, Byrnes and Hall (see also \citep{Niemeyer:1997mt,Yokoyama:1998xd,Musco:2008hv} for pioneer work) revisited PBH mass functions from peaked primordial spectra using an updated treatment of the collapse mapping, including critical behavior, and showed that the resulting profiles can differ significantly from simple lognormal templates \cite{Gow:2020cou}. In the narrow-spectrum limit, this behavior reflects the fact that the critical mass map itself provides an irreducible source of width, rather than the mass function being determined only by the detailed shape of the primordial peak. Our calculation can be viewed as the DSS extension of thi {standard power-law} mass-function envelope: the minimum-width profile associated with the {standard power-law} critical-scaling relation is now modulated by the log-periodic residual of the massless-scalar critical solution. The new question is whether this DSS substructure remains visible once the mass map is folded with a realistic primordial spectrum and {window function}.

The second relevant context is the work of Yoo, Harada, Hirano and Kohri, who developed a peak-theory procedure for arbitrary power spectra and emphasized that the {choice of window function} matters, particularly for broad spectra \cite{Yoo:2020dkz}. We therefore treat the {dependence on the choice of window function} as part of the problem rather than as a cosmetic plotting choice.

We find that the {standard power-law} critical-scaling relation maps the continuous distribution of supercritical perturbation amplitudes into an irreducible minimum width in the PBH mass function, while DSS modulates the resulting {standard power-law} mass-function envelope with approximately log-periodic substructure. In the narrow-spectrum limit, the characteristic separation between adjacent DSS-induced substructures is controlled by the DSS period and corresponds to a multiplicative mass spacing of about $5.6$ for the fiducial Choptuik parameters used below. For sufficiently broad primordial spectra, however, the convolution over horizon masses dephases the DSS modulation, and the mass function becomes controlled primarily by the width of the primordial spectrum rather than by the near-threshold critical structure.

The paper is organized as follows. In Sec.~\ref{sec:critical_collapse} we introduce the DSS-modulated critical mass-scaling law and discuss the minimum width generated by critical behavior, together with the characteristic DSS mass spacing. In Sec.~\ref{sec:mass_function_formalism} we describe the PBH mass-function framework and numerical implementation, including the primordial curvature spectrum, {window functions}, peak weighting, horizon-mass mapping, kination-era evolution, and the peak-alignment procedure used to compare profile shapes. In Sec.~\ref{sec:results} we present the narrow-spectrum and broad-spectrum comparisons and introduce residual and periodogram diagnostics to quantify the coherence and progressive washout of the DSS signal. In Sec.~\ref{sec:observational_implications} we discuss possible observational implications, and in Sec.~\ref{sec:conclusions} we summarize our conclusions. Appendix~\ref{app:tophat_domain_sensitivity} examines the integration-domain sensitivity of the real-space top-hat window function.

% ============================================================
\section{DSS-modulated critical behavior}
\label{sec:critical_collapse}

\subsection{Critical scaling and DSS fitting function}
\label{subsec:dss_template}

In Ref.~\cite{Padilla:2026dss}, each family of long-wavelength initial data
was labeled by the central curvature amplitude $p$. The corresponding
DSS-modulated critical mass-scaling relation was therefore obtained as a
function of the logarithmic distance to threshold,
$\ln(p-p_{\rm th})$. In the present work, we rewrite this relation in terms
of a density amplitude more directly connected to the statistical
mass-function calculation.

For a fixed initial-data family, the curvature profile on the initial
comoving hypersurface can be written as
\begin{equation}
K(r)=pF(r),
\qquad
F(0)=1.
\end{equation}
At leading order in the long-wavelength expansion, the compaction function is
related to the curvature profile by
\begin{equation}
\mathcal{C}(r)=A_w K(r)r^2,
\end{equation}
where $A_w$ depends on the equation of state and on the convention adopted
for the compaction{~\cite{Harada:2015yda}}. Since changing $p$ rescales the profile without changing
its shape, the position $r_m$ of the maximum compaction is independent of
$p$ within a fixed family, and
\begin{equation}
\mathcal{C}_{\rm max}=q_F p,
\label{eq:Cmax_p_relation}
\end{equation}
where $q_F$ is a family-dependent constant.

The collapse thresholds in Ref.~\cite{Padilla:2026dss} were also reported in
terms of $\mathcal{C}_{\rm max}$. In comoving slicing, the compaction and the
volume-averaged density contrast are related, at leading order in the
gradient expansion, by
\begin{equation}
\mathcal{C}(t,r)
=
\left[H(t)R_A(t,r)\right]^2
\bar{\delta}(t,r),
\label{eq:compaction_density_relation}
\end{equation}
where $R_A(t,r)$ is the areal radius. Extrapolating the long-wavelength
solution to horizon entry of the characteristic scale, $H_{\rm HC}R_A(t_{\rm HC},r_m)=1$, gives
\begin{equation}
\bar{\delta}_{\rm HC}(r_m)
=
\mathcal{C}_{\rm max}.
\label{eq:deltaH_compaction_relation}
\end{equation}
For the long-wavelength initial-data families considered in
Ref.~\cite{Padilla:2026dss}, we define
\begin{equation}
\delta_H
\equiv
\bar{\delta}_{\rm HC}(r_m).
\label{eq:deltaH_definition}
\end{equation}
Within this extrapolated long-wavelength convention, the corresponding
threshold satisfies{~\cite{Harada:2015yda}}
\begin{equation}
\delta_{H,\rm th}
=
\mathcal{C}_{{\rm max},{\rm th}}.
\label{eq:deltaH_threshold_relation}
\end{equation}
Equations~\eqref{eq:Cmax_p_relation} and
\eqref{eq:deltaH_compaction_relation} then imply, for each fixed
initial-data family,
\begin{equation}
\delta_H=q_Fp,
\qquad
\delta_{H,\rm th}=q_Fp_{\rm th}.
\end{equation}
Consequently,
\begin{equation}
\ln\left(\delta_H-\delta_{H,\rm th}\right)
=
\ln\left(p-p_{\rm th}\right)
+
\ln q_F.
\label{eq:p_deltaH_log_relation}
\end{equation}

Thus, replacing the central curvature amplitude by the extrapolated
volume-averaged density amplitude produces only a constant shift in the
logarithmic distance to threshold. This shift can be absorbed into the
normalization of the mass-scaling relation and into the phase of the DSS
residual. It does not change the critical exponent, the logarithmic period,
or the characteristic DSS mass spacing.

We therefore introduce
\begin{equation}
x
\equiv
\ln\left(\delta_H-\delta_{H,\rm th}\right),
\label{eq:x_deltaH_def}
\end{equation}
and write the DSS-modulated critical mass-scaling relation as
\begin{equation}
\ln M_{\rm PBH}
=
\ln\left(\mathcal{K}M_{\rm H}\right)
+
\gamma x
+
f_{\rm DSS}(x).
\label{eq:dss_scaling}
\end{equation}
The {standard power-law} critical-scaling relation in
Eq.~\eqref{eq:intro_standard_scaling} is recovered for
$f_{\rm DSS}=0$.

In the mass-function calculation below, the same symbol $\delta_H$ is
used for the effective smoothed-density amplitude
associated with this collapse variable. The precise tuning of the threshold in terms of the statistical variable is important for predictions
of the absolute PBH abundance, whereas the present work focuses primarily on
normalized profiles and on the propagation of the DSS modulation.

We model the residual using the fitting template introduced in Ref.~\cite{Padilla:2026dss},
\begin{equation}
f_{\rm DSS}(x)
=C_0+
\Adss\ln\left[
\eps+
\left|
\cos\left(
\frac{\pi x}{\Pln}+\phi
\right)
\right|
\right].
\label{eq:dss_model}
\end{equation}
The DSS residual is a physical consequence of the discretely self-similar critical solution rather than a fitting artifact. Equation~\eqref{eq:dss_model} is a fitting function of the residual pattern observed in the numerical-relativity simulations of Ref.~\cite{Padilla:2026dss}. We use it here to propagate the cosmological DSS modulation into the PBH mass-function calculation.

The fiducial parameters used in the numerical examples are
\begin{subequations}
\begin{equation}
  \mathcal{K}=10,
  \qquad
  \gamma=\gamma_{\rm Ch}=0.374,
  \qquad
  \deltaHth=0.547,
  \label{eq:standard_params}
\end{equation}
for the {standard power-law} critical-scaling relation, and
\begin{eqnarray}
  \Pln=\Pln^{\rm Ch}=4.60,&&
  \qquad
  \Adss=0.30,\nonumber\\
  \qquad
  \eps=10^{-6},&&
  \qquad
  \phi=C_0=0,
  \label{eq:dss_params}
\end{eqnarray}
\end{subequations}
for the DSS modulation. The threshold $\deltaHth=0.547$ is the average of the thresholds obtained for the two initial-data families in Ref.~\cite{Padilla:2026dss}. The scaling exponent and DSS period are chosen close to the standard Choptuik massless-scalar values, consistently with the cosmological match found in Ref.~\cite{Padilla:2026dss}. The amplitude $\Adss\simeq0.30$ is a parameter controlling the visibility of the residuals and is chosen to be consistent with the modulation amplitude found in Ref.~\cite{Padilla:2026dss}.

Because of the absolute value in Eq.~\eqref{eq:dss_model}, the modulation is periodic under $x\rightarrow x+\Pln$, so that $f_{\rm DSS}(x+\Pln)=f_{\rm DSS}(x)$. The effective period in the logarithmic distance to threshold is therefore $\Pln$. This is the period that is subsequently mapped into the PBH mass function in the narrow-spectrum limit. The physical role of Eq.~\eqref{eq:dss_model} is simple: {standard power-law} critical-scaling relation maps equal intervals in $x$ to equal intervals in $\ln \MBH$, while the DSS residual periodically stretches and compresses this mapping and, when the modulation is sufficiently pronounced, can also make it locally non-monotonic. In that case, different distances from threshold{, $\delta_{\rm H}-\delta_{\rm H, th}$,} can contribute to the same PBH mass, so the mass function receives contributions from multiple branches of the critical scaling relation. Since the mass function involves the inverse Jacobian between $x$ and $\ln \MBH$,
\begin{equation}
\frac{\dd \ln \MBH}{\dd x}=
\gamma+f_{\rm DSS}'(x),
\label{eq:dss_jacobian}
\end{equation}
regions where this derivative becomes small produce localized enhancements in mass space. The detailed amplitude and shape of these features are model dependent, but their approximate logarithmic spacing is controlled mainly by $\gamma\Pln$, as discussed below.

\begin{figure}[htbp]
\centering
\maybeincludegraphics[width=0.95\linewidth]{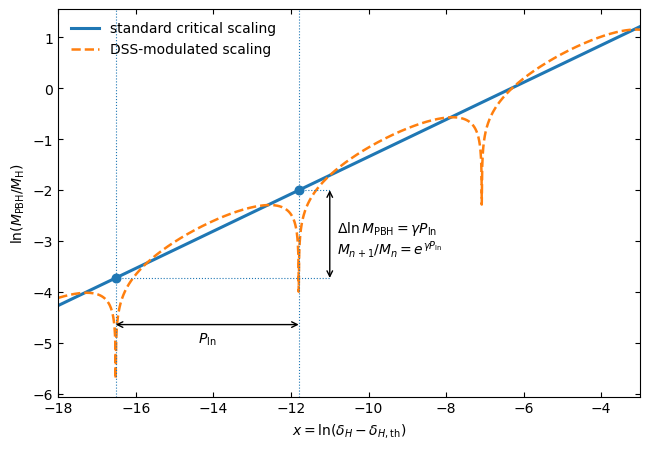}
\caption{\footnotesize{
Schematic DSS-modulated critical scaling relation. The solid line shows the {standard power-law} critical-scaling relation $\ln\MBH=\ln(\mathcal{K}\MH)+\gamma x$, while the dashed line includes the DSS correction $f_{\rm DSS}(x)$. The vertical dotted lines mark two consecutive minima of the DSS modulation, separated by one period $P_{\ln}$ in $x=\ln(\delta_H-\delta_{H,\rm th})$. The blue points show the corresponding locations on the {standard power-law} critical-scaling relation. Since this relation is linear in $x$, this shift gives $\Delta\ln\MBH=\gamma P_{\ln}$, or equivalently $\MBH^{(n+1)}/\MBH^{(n)}=\exp(\gamma P_{\ln})$. This exact fixed-horizon-mass spacing in the critical map provides the reference scale $\Delta\ln m\simeq\gamma P_{\ln}$ for structures in the final mass function. The plotted modulation is a fitting parametrization of the residual pattern observed in Ref.~\cite{Padilla:2026dss}.
}}
\label{fig:scaling_relation}

\end{figure}

Figure~\ref{fig:scaling_relation} illustrates this interpretation. The DSS term produces a log-periodic modulation of the otherwise linear relation between $x$ and $\ln \MBH$, leading to alternating regions where the mass map is stretched, compressed, and, for sufficiently large modulation, can become locally non-monotonic.

% ============================================================
\subsection{Minimum width and DSS mass spacing}
\label{sec:minimum_width}
The first qualitative effect is independent of DSS. Even if the primordial power spectrum is infinitely narrow, the PBH mass distribution has a finite width because the {standard power-law} critical-scaling relation takes the continuous distribution of supercritical amplitudes into a continuous distribution of PBH masses. To see this explicitly, consider the fixed-horizon-mass limit. A Dirac-delta function primordial spectrum fixes the characteristic horizon mass
$\MH$, but it does not fix the smoothed density amplitude above threshold. Using $f_{\rm DSS} = 0$ in Eq.~\eqref{eq:dss_scaling}, we obtain
\begin{equation}
  \ln \MBH = \ln(\mathcal{K}\MH)+\gamma x .
  \label{eq:fixed_horizon_standard}
\end{equation}
Therefore a continuous probability distribution for $x$ immediately becomes
a continuous probability distribution for $\ln \MBH$. The low-mass side
corresponds to perturbations very close to threshold, $x\to-\infty$, while
the high-mass side corresponds to rarer, larger-amplitude perturbations.
This is the irreducible minimum width generated by critical behavior. In this
sense, the resulting {standard power-law} mass function should be interpreted as a smooth
mass-function envelope whose existence does not rely on a finite primordial width. This
point is closely related to the minimum-width result of
Ref.~\cite{Gow:2020cou}, which showed that narrow peaks in the primordial
spectrum lead to PBH mass distributions controlled mainly by critical
behavior rather than by the detailed shape of the peak.

The DSS modulation does not create the width from scratch. Instead, it
modulates this minimum-width profile. To estimate the characteristic spacing,
we first evaluate Eq.~\eqref{eq:dss_scaling} at fixed $\MH$. Since
$f_{\rm DSS}(x)$ is periodic with period $\Pln$, two points separated by one
full DSS period and corresponding to the same phase of the modulation satisfy
\begin{align}
\ln \MBH(x+\Pln)-\ln \MBH(x)
&=\gamma\Pln .
\label{eq:spacing_derivation}
\end{align}
This relation is exact at fixed horizon mass. It therefore motivates the leading expectation
\begin{equation}
\Delta\ln m \simeq \gamma\Pln
\label{eq:dss_spacing_general}
\end{equation}
for the separation between adjacent DSS-induced structures in the full PBH mass function. Equivalently, the corresponding multiplicative spacing between their locations in mass space is
\begin{equation}
\frac{m_{n+1}}{m_n}
\simeq
\exp(\gamma\Pln).
\label{eq:mass_spacing_general}
\end{equation}
Figure~\ref{fig:scaling_relation} illustrates the origin of this estimate: moving from one oscillation of a given DSS phase to the next corresponds to a shift $\Pln$ in $x$, which the {standard power-law} critical-scaling relation maps into a shift $\gamma\Pln$ in $\ln\MBH$.

For the fiducial Choptuik DSS parameters in Eqs.~\eqref{eq:standard_params} and \eqref{eq:dss_params}, the corresponding expectation for the final mass function is
\begin{align}
\Delta\ln m \simeq1.72,
\qquad
\frac{m_{n+1}}{m_n}
\simeq5.6 .
\label{eq:dss_spacing_number}
\end{align}
This is the origin of the characteristic factor $\sim5.6$ quoted in the abstract and introduction.

The exact relation in Eq.~\eqref{eq:spacing_derivation} concerns the critical mass map in the idealized fixed-horizon-mass limit, where two points at the same DSS phase can be compared directly. The relations in Eqs.~\eqref{eq:dss_spacing_general} and \eqref{eq:mass_spacing_general} instead refer to the locations of structures in the final mass function. In the full calculation, the integration over smoothing scales superposes different horizon masses, while a finite-width primordial spectrum contributes over a range of scales. The visible maxima, minima, or shoulders in the final mass distribution can therefore be shifted and their contrast reduced. Nevertheless, $\exp(\gamma\Pln)$ remains the natural reference scale for identifying the DSS pattern in the narrow-spectrum limit.

% ============================================================
\section{Mass-function framework and numerical strategy}
\label{sec:mass_function_formalism}

\subsection{Normalized mass function and schematic construction}
\label{subsec:mass_function_definition}

We define the normalized PBH mass function $\psi(m)$ by
\begin{equation}
  \int_0^\infty \psi(m)\,\dd m = 1 .
  \label{eq:psi_norm}
\end{equation}
With this convention, the normalized distribution per logarithmic PBH mass interval is
\begin{equation}
  \frac{\dd P}{\dd\ln m}
  =
  m\psi(m),
  \label{eq:mpsi_def}
\end{equation}
where $P$ denotes the probability that a PBH in the population has
mass in a given range. Thus $m\psi(m)$ is the normalized distribution per
logarithmic mass interval.

\subsection{Primordial spectrum, smoothing, and peak weighting}
\label{subsec:ingredients}

\paragraph{Primordial spectrum and density variance.}
\label{subsec:primordial_spectrum}

We take the primordial curvature power spectrum to be lognormal,
\begin{equation}
  \Pzeta(k)=\frac{A}{\sqrt{2\pi}\Delta}
  \exp\left[-\frac{1}{2}\left(\frac{\ln(k/k_p)}{\Delta}\right)^2\right],
  \label{eq:lognormal_pzeta}
\end{equation}
where $A$ controls the integrated power, $k_p$ is the characteristic scale, and
$\Delta$ is the logarithmic width. In the distributional limit $\Delta\to0$, the lognormal spectrum approaches
a Dirac-delta spectrum. In the calculations below, however, the Dirac-delta
case is evaluated separately as a normal Dirac-delta function rather than by setting $\Delta=0$ in
Eq.~\eqref{eq:lognormal_pzeta}. 

For a constant equation of state $w$, we use the linearized superhorizon relation \citep{Young:2019yug}
\begin{equation}
\delta(k,t)=C_w\left(\frac{k}{aH}\right)^2\zeta(k),
\qquad
C_w=\frac{2(1+w)}{5+3w},
\label{eq:delta_zeta_relation}
\end{equation}
so that, evaluating each smoothing scale $R$ at its corresponding horizon-entry time, $aH=R^{-1}$, the smoothed density power spectrum is
\begin{equation}
\Pdelta(k,R)=C_w^2(kR)^4W^2(kR)\Pzeta(k).
\label{eq:pdelta_smoothed}
\end{equation}
The spectral moments are then
\begin{equation}
  \sigma_n^2(R)=\int \dd\ln k\,k^{2n}\Pdelta(k,R).
  \label{eq:sigma_moments}
\end{equation}
In the numerical examples below we focus on a kination era in the
timelike-gradient regime discussed in the Introduction. We therefore set
$w=1$, for which $C_w=1/2$.

\paragraph{Window functions.}
\label{subsec:smoothing_prescriptions} We consider {the following} four {window functions:
the Gaussian, the $k$-space top-hat ($k$TH), 
the real-space top-hat (RTH), and 
the real-space top-hat regulated by the factor of the transfer function for a stiff matter (TRTH), given by} 
{\begin{subequations}
\label{eq:window_functions}
\begin{alignat}{2}
&\mbox{Gaussian}: &W_{\rm G}(z)=\exp(-z^2/2),
\label{eq:gaussian_window}\\
&\mbox{$k$TH}: &W_{k{\rm TH}}(z)=\Theta(1-z),
\label{eq:kspace_window}\\
&\mbox{RTH}: &W_{\rm TH}(z)=3\frac{\sin z-z\cos z}{z^3},
\label{eq:tophat_window}\\
&\mbox{TRTH}: &W_{{\rm TH}+T}(z)=W_{\rm TH}(z)T_{\rm stiff}(z),
\label{eq:tophat_transfer_window}
\end{alignat}
\end{subequations}}
where $z=kR$. The transfer function used in the last expression is
\begin{equation}
T_{\rm stiff}(z)=\frac{2J_1(z/2)}{z/2}\,\,,
\label{eq:stiff_transfer}
\end{equation}
where $J_1$ denotes the Bessel function of the first kind of order one. Its small-$z$ limit satisfies $T_{\rm stiff}(z)=1+\mathcal{O}(z^2)$, so the
large-scale normalization is unchanged.

The window function $W$ specifies how Fourier modes in the primordial curvature
spectrum are combined to define the perturbation amplitude on a finite
smoothing scale $R$. Since PBH formation is controlled by the collapse of
finite regions rather than by individual Fourier modes, this projection is an
unavoidable part of the mass-function calculation. The Gaussian {and $k$TH window functions} provide two standard Fourier-space {window function}s, while
the {RTH} window function corresponds to a spherical average of the density
contrast in real space.

We also include the {TRTH window function}. This {window function} retains the
spherical-average interpretation of the RTH window function while {perhaps potentially} accounting for
the linear subhorizon evolution of perturbations around a $w=1$ background.
It also improves the large-$z$ behavior of the effective window function, where the {RTH} window function has an oscillatory and slowly decaying Fourier-space
tail.

This last point is important for the interpretation of the numerical results.
The RTH window function is not treated as a uniquely preferred smoothing
prescription in this work. Its Fourier-space representation oscillates and
decays only as a power law, so the associated spectral moments {\eqref{eq:sigma_moments}} 
can be
especially sensitive to the finite integration domain, the numerical
resolution, and the tail treatment. We therefore keep the RTH window function in the comparison as a diagnostic of smoothing sensitivity, but we
interpret detailed features in that panel with caution. 
The TRTH window function provides a more controlled behaviour for
the kination case because it damps the high-$kR$ oscillatory tail.

The purpose of comparing {these window functions} is not to determine a unique
optimal window function, but to use the {window-function dependence} as a robustness check. We therefore focus on features that persist across the comparison {over the window functions}, in particular the minimum width induced by critical scaling, the approximate DSS spacing
$\Delta\ln m\simeq\gamma\Pln$ in the final mass function, and the washout of DSS substructure for broad primordial spectra. Conversely, features that appear only in the RTH panel should be regarded as sensitive to the choice of window function and numerical domain, rather than as robust predictions of DSS.

\paragraph{Peak weighting and horizon-mass mapping.}
\label{subsec:peak_weighting}

We estimate the relative abundance of rare, high-amplitude configurations
using the traditional peak-theory prescription adopted in
Ref.~\cite{Gow:2020cou}. In the high-peak approximation, the differential
peak number density is
\begin{equation}
  n_{\rm pk}(\nu,R)=
  \frac{1}{3^{3/2}(2\pi)^2}
  \left(\frac{\sigma_1}{\sigma_0}\right)^3
  \nu^3\exp\left(-\frac{\nu^2}{2}\right),
  \label{eq:peak_number_density}
\end{equation}
where
\begin{equation}
  \nu=\frac{\delta_H}{\sigma_0(R)}.
\end{equation}
Here $\delta_H$ is the effective smoothed-density amplitude introduced in
Sec.~\ref{subsec:dss_template}, extrapolated to horizon entry. Its
spectral moments are obtained from Eq.~\eqref{eq:sigma_moments}, using the
linearized relation between the primordial curvature perturbation and the
density contrast. For Gaussian primordial curvature perturbations, the
smoothed linear-density field is therefore also Gaussian, with variance
$\sigma_0^2(R)$.

Operationally, we use the same effective density amplitude in the peak
weighting and in the critical mass relation. For the
RTH window, the smoothed density has the direct interpretation
of a volume-averaged density contrast. For the other window functions,
$\delta_H$ should instead be understood as the corresponding filtered density
amplitude.

The collapse threshold entering the statistical weighting should, strictly
speaking, be tuned in terms of the same Gaussian smoothed-density
amplitude whose variance is $\sigma_0^2(R)$. A threshold obtained instead
from a nonlinear volume-averaged density contrast or from the maximum
compaction does not automatically coincide with the threshold of the
smoothed linear-density field. Establishing this correspondence is important
for predictions of the absolute PBH abundance, owing to its exponential
sensitivity to the threshold.

In the present simplified treatment, we adopt $\delta_{H,\rm th}$ as an
effective threshold in the smoothed-density parametrization and focus on
normalized mass-function profiles. The main purpose of the peak weighting is
therefore to describe the relative contribution of rare supercritical
configurations and to study how the DSS-modulated critical mass map is
propagated through the mass-function calculation. A fully quantitative
prediction of the absolute abundance would require a consistent determination
of the collapse threshold and critical mass relation in terms of the same
smoothed-density variable used in the peak statistics. This uncertainty does
not affect the existence of the DSS modulation in the critical mass map,
although it can affect its normalization, phase, and detailed visibility in
the final mass function.
% % ============================================================
\subsection{Construction of the mass function}
\label{subsec:mass_function_construction}

We now combine the ingredients introduced above to construct the PBH mass function following Ref.~\citep{Gow:2020bzo}. We use $m$ as the continuous mass variable entering $\psi(m)$, while $M_{\rm PBH}(x,R)$ denotes the PBH mass assigned by the critical mass map to a perturbation characterized by $x$ and the smoothing scale $R$.

For a constant background equation of state $w$, the horizon mass associated with the smoothing scale $R$ is
\begin{equation}
M_{\rm H}(R)
=M_{{\rm H},{\rm ref}}
\left(\frac{R}{R_{\rm ref}}\right)^q,
\qquad
q=\frac{3(1+w)}{1+3w},
\label{eq:horizon_mass_R}
\end{equation}
where $R_{\rm ref}$ is a reference smoothing scale and
$M_{{\rm H},{\rm ref}}\equiv M_{\rm H}(R_{\rm ref})$. Equivalently, defining
$k_{\rm ref}\equiv R_{\rm ref}^{-1}$, this relation can be written as
\begin{equation}
M_{\rm H}(R)
=M_{{\rm H},{\rm ref}}
(k_{\rm ref}R)^q.
\label{eq:horizon_mass_kref}
\end{equation}
For kination, $w=1$ and therefore $q=3/2$.

Using the distance-to-threshold variable defined in
Eq.~\eqref{eq:x_deltaH_def}, so that
$\delta_H=\delta_{H,\rm th}+e^x$, the PBH mass associated with a given pair
$(x,R)$ is
\begin{equation}
M_{\rm PBH}(x,R)
=\mathcal{K} M_{\rm H}(R)
\exp\left[\gamma x+f_{\rm DSS}(x)\right].
\label{eq:mass_map_Rx}
\end{equation}

The peak height entering the high-peak abundance is
\begin{equation}
\nu(x,R)
=\frac{\delta_{H,\rm th}+e^x}{\sigma_0(R)}.
\label{eq:nu_x_R}
\end{equation}
At a fixed smoothing scale, the corresponding PBH energy fraction at formation is
\begin{equation}
\beta(R)
\propto R^3
\int_{\delta_{H,\rm th}}^\infty
\dd\delta_H
\frac{M_{\rm PBH}(\delta_H,R)}
{M_{\rm H}(R)}
n_{\rm pk}
\left[
\frac{\delta_H}{\sigma_0(R)},R
\right].
\label{eq:beta_peaks}
\end{equation}
Here the factor $M_{\rm PBH}/M_{\rm H}$ converts the number density of collapsing peaks into the associated PBH energy density fraction, while the factor $R^3$ follows the traditional peak-theory prescription of Ref.~\citep{Gow:2020bzo}, in which $n_{\rm pk}$ is a peak number density per unit physical volume.

We assume that the PBHs considered here form during a kination era that ends when the comoving horizon scale reaches $R_{\rm k\rightarrow r}$. During kination, the dominant scalar-field background redshifts as $a^{-6}$, whereas the PBH energy density redshifts as $a^{-3}$. The PBH fraction therefore grows as $a^3$. Since
$R=(aH)^{-1}\propto a^2$ for $w=1$, the growth between PBH formation at scale $R$ and the end of kination is
\begin{equation}
\frac{\beta(R_{\rm k\rightarrow r})}{\beta(R)}
=\left(\frac{R_{\rm k\rightarrow r}}{R}\right)^{3/2}.
\label{eq:kination_growth_factor}
\end{equation}
After the transition to radiation domination, the PBH fraction grows linearly with the scale factor. The additional growth between the end of kination and matter--radiation equality is therefore $R_{\rm eq}/R_{\rm k\rightarrow r}$. The abundance at equality is consequently
\begin{equation}
\Omega_{\rm PBH}
\propto
\int_0^{R_{\rm k\rightarrow r}}
\dd\ln R
\left(\frac{R_{\rm k\rightarrow r}}{R}\right)^{3/2}
\left(\frac{R_{\rm eq}}{R_{\rm k\rightarrow r}}\right)
\beta(R).
\label{eq:Omega_integral_R}
\end{equation}
The factors depending only on $R_{\rm k\rightarrow r}$ and $R_{\rm eq}$ affect the overall abundance but cancel from the normalized mass function. The scale-dependent factor $R^{-3/2}$ must nevertheless be retained because it changes the relative weighting of PBHs formed at different horizon scales.

To resolve the total PBH abundance into the final PBH mass, we insert
\begin{equation}
1=
\int \dd\ln m\,
\delta_{\rm D}
\left[
\ln m-\ln M_{\rm PBH}(x,R)
\right],
\end{equation}
which assigns each contribution labelled by $(x,R)$ to its corresponding PBH mass
$m=M_{\rm PBH}(x,R)$. Comparing the resulting expression with
\begin{equation}
\Omega_{\rm PBH}
=
\int \dd\ln m\,
\frac{\dd\Omega_{\rm PBH}}{\dd\ln m},
\end{equation}
and combining the $R^3$ factor in Eq.~\eqref{eq:beta_peaks} with the
kination growth factor $R^{-3/2}$, while omitting overall constants
independent of $R$, we obtain
\begin{widetext}
\begin{equation}
\begin{aligned}
\frac{\dd\Omega_{\rm PBH}}{\dd\ln m}
\propto{}&
\int_0^{R_{\rm k\rightarrow r}} \dd\ln R
\int_{-\infty}^\infty \dd x
R^{3/2}
\frac{M_{\rm PBH}(x,R)}
{M_{\rm H}(R)}
n_{\rm pk}\left[\nu(x,R),R\right]
e^x\times
\delta_{\rm D}
\left[
\ln m-\ln M_{\rm PBH}(x,R)
\right].
\end{aligned}
\label{eq:mass_function_integral}
\end{equation}
\end{widetext}
The factor $e^x$ is the Jacobian associated with
$\dd\delta_H=e^x\dd x$. The Dirac-delta function also accounts automatically for multiple branches if the DSS-modulated mass map becomes locally non-monotonic.

The corresponding unnormalized mass function is
\begin{equation}
\widetilde{\psi}(m)
=\frac{1}{m}
\frac{\dd\Omega_{\rm PBH}}{\dd\ln m}.
\label{eq:psi_unnormalized}
\end{equation}
The normalized mass function $\psi(m)$ is then obtained using
Eq.~\eqref{eq:psi_norm}. The numerical calculation consists of the wavenumber integral entering the spectral moments, followed by the integrations over the smoothing scale $R$ and the supercritical amplitude $x$. The formally infinite domains in $k$ and $x$ are replaced by finite ranges chosen to contain the relevant support of the integrands, while the smoothing-scale integration is restricted to $R<R_{\rm k\rightarrow r}$ so that all included PBHs form during kination.

% % ============================================================
\subsection{Peak alignment and numerical strategy}
\label{sec:numerical_strategy}

For all finite-width spectra considered below, we keep the maximum height of the primordial curvature spectrum fixed,
\begin{equation}
\Pzeta^{\rm peak}=0.05.
\label{eq:fixed_Pzeta_peak}
\end{equation}
With the normalization used in Eq.~\eqref{eq:lognormal_pzeta}, this implies
\begin{equation}
A(\Delta)
=\sqrt{2\pi}\Delta\Pzeta^{\rm peak},
\qquad \Delta>0.
\label{eq:A_of_Delta_fixed_peak}
\end{equation}
The parameter $A$ is therefore not the peak height itself, but the power integrated over $\dd\ln k$. Its dependence on $\Delta$ in Eq.~\eqref{eq:A_of_Delta_fixed_peak} is precisely what is required to keep $\Pzeta^{\rm peak}$ unchanged as the spectrum is broadened. The Dirac-delta reference spectrum is evaluated separately rather than as
$\Delta=0$ in Eq.~\eqref{eq:lognormal_pzeta}. Its integrated amplitude is
chosen to equal that of the narrowest finite-width spectrum,
$\Delta=0.05$,
\begin{equation}
A_\delta
=
\sqrt{2\pi}(0.05)\Pzeta^{\rm peak}
\simeq 6.27\times10^{-3}.
\end{equation}

As noted in Ref.~\cite{Gow:2020cou}, broadening the primordial spectrum also shifts the resulting PBH mass distribution. To facilitate a direct comparison of the profile shapes, for each primordial width $\Delta$ and window function $W$ we adjust only the primordial-spectrum peak position $k_p$. We denote the resulting aligned value by $k_p(\Delta;W)$. For each window function separately, $k_p(\Delta;W)$ is chosen such that the peak of the standard, non-DSS-modulated distribution $m\psi_{\rm std}(m)$ coincides with that of the corresponding Dirac-delta reference case computed using the same window function. The same value of $k_p(\Delta;W)$ is then used for the corresponding DSS-modulated curve.

Changing $k_p$ changes the characteristic horizon mass according to $M_{\rm H}(k_p)\propto k_p^{-q}$ {from Eq.~\eqref{eq:horizon_mass_R}}. Since the PBH mass is proportional to the horizon mass up to the critical-scaling factor, varying $k_p$ mainly shifts the mass function horizontally in $\ln m$. This alignment therefore removes the overall change in the characteristic mass scale and allows changes in the profile width and DSS substructure to be compared more directly.

We study two ranges of primordial widths. The narrow-spectrum comparison uses
\begin{equation}
\Delta=0,0.05,0.1,0.2,0.3,
\label{eq:narrow_width_values}
\end{equation}
and is designed to illustrate the critical-scaling-dominated regime{, where and hereafter the case of} $\Delta=0$ denotes the separate Dirac-delta reference case {with the same integrated amplitude as for $\Delta =0.05$ as we already discussed.} The broad-width comparison uses
\begin{equation}
\Delta=0,0.5,1,2,3.
\label{eq:width_scan_values}
\end{equation}

We define
\begin{equation}
k_{\rm k\rightarrow r}\equiv R_{\rm k\rightarrow r}^{-1}
\end{equation}
as the comoving horizon scale at the end of the kination era. Modes with $k>k_{\rm k\rightarrow r}$ enter the horizon during kination. In the numerical examples we take
\begin{equation}
k_{\rm k\rightarrow r}=10^8~{\rm Mpc}^{-1},
\qquad
R_{\rm k\rightarrow r}=10^{-8}~{\rm Mpc}.
\label{eq:kination_end_scale}
\end{equation}
This value corresponds to a transition well before Big Bang nucleosynthesis. In particular, successful BBN requires the transition from kination to radiation domination to occur at temperatures above a few MeV, with the conservative condition $T_{\rm k\rightarrow r}\gtrsim4~{\rm MeV}$ \citep{Maki:2026qpf}. The corresponding horizon scale around the BBN epoch is of order $k_{\rm BBN}\sim10^4~{\rm Mpc}^{-1}$ \citep{Inomata:2016uip}. Our fiducial choice $k_{\rm k\rightarrow r}=10^8~{\rm Mpc}^{-1}$ therefore places the end of kination safely before BBN. It is also sufficiently smaller than the wavenumbers carrying significant power in all the spectra considered below, so that the relevant PBHs form during kination.

For the kination-era horizon-mass relation, we choose the reference scale to be the end of kination,
\begin{equation}
R_{\rm ref}=R_{\rm k\rightarrow r},
\qquad
k_{\rm ref}=k_{\rm k\rightarrow r}.
\label{eq:kin_reference_scale}
\end{equation}
The corresponding reference horizon mass is obtained from the radiation-dominated relation after the transition, \begin{eqnarray}
M_{\rm H}(k_{\rm k\rightarrow r}){\simeq1.8\times10^{-3}~M_\odot
\left(
\frac{k_{\rm k\rightarrow r}}
{10^8~{\rm Mpc}^{-1}}
\right)^{-2}}
.
\label{eq:horizon_mass_at_kination_end}
\end{eqnarray}
For modes entering the horizon during kination, the horizon mass is therefore
\begin{equation}
M_{\rm H}(k)
=M_{\rm H}(k_{\rm k\rightarrow r})
\left(
\frac{k}{k_{\rm k\rightarrow r}}
\right)^{-3/2},
\qquad
k>k_{\rm k\rightarrow r}.
\label{eq:kination_horizon_mass_scaling}
\end{equation}

The fiducial mass scale is chosen in the asteroid-mass range. The target horizon mass
\begin{equation}
M_{\rm H}(k_p)
\simeq10^{-15}~M_\odot
\simeq1.99\times10^{18}~{\rm g}
\label{eq:asteroid_mass_target}
\end{equation}
corresponds to
\begin{equation}
k_p
\simeq1.5\times10^{16}~{\rm Mpc}^{-1}
\label{eq:asteroid_peak_scale}
\end{equation}
for the fiducial value $k_{\rm k\rightarrow r}{=10^{8}{\rm Mpc}^{-1}}$ adopted above. This value sets the horizontal mass scale of the Dirac-delta function reference case.

The numerical integrations are performed over finite ranges in
$k$, $R$, $x$, and $m$. For all mass-function curves shown in the
main text, the smoothing-scale domain is taken to be
\begin{equation}
  \frac{3\times10^{-5}}{k_p(\Delta;W)}
  \leq R \leq
  \min\left[
    \frac{3\times10^{5}}{k_p(\Delta;W)},
    R_{\rm k\rightarrow r}
  \right].
  \label{eq:fiducial_R_domain}
\end{equation}
The upper bound $R_{\rm k\rightarrow r}$ ensures that
all included scales enter the horizon during the kination
era. For the values of $k_p(\Delta;W)$ relevant to the main figures, the
fiducial upper limit $3\times10^{5}/k_p(\Delta;W)$ generally lies below
$R_{\rm k\rightarrow r}$, so the results probe only a subset of the full
kination-entry domain.

The Gaussian window function and $k$TH window function as well as the TRTH results are stable when this range is enlarged.
The RTH result is more sensitive to the upper
smoothing-scale boundary because its effective large-$kR$ behavior is
not strongly damped. Appendix~\ref{app:tophat_domain_sensitivity}
compares different $R$ domains and shows explicitly which features
are stable and which track the adopted cutoff.

% ============================================================
\section{Results}
\label{sec:results}

% ============================================================
\subsection{Narrow spectra and the persistence of the critical width}
\label{sec:narrow_results}

\begin{figure*}[htbp]
\centering
\maybeincludegraphics[width=0.98\textwidth]{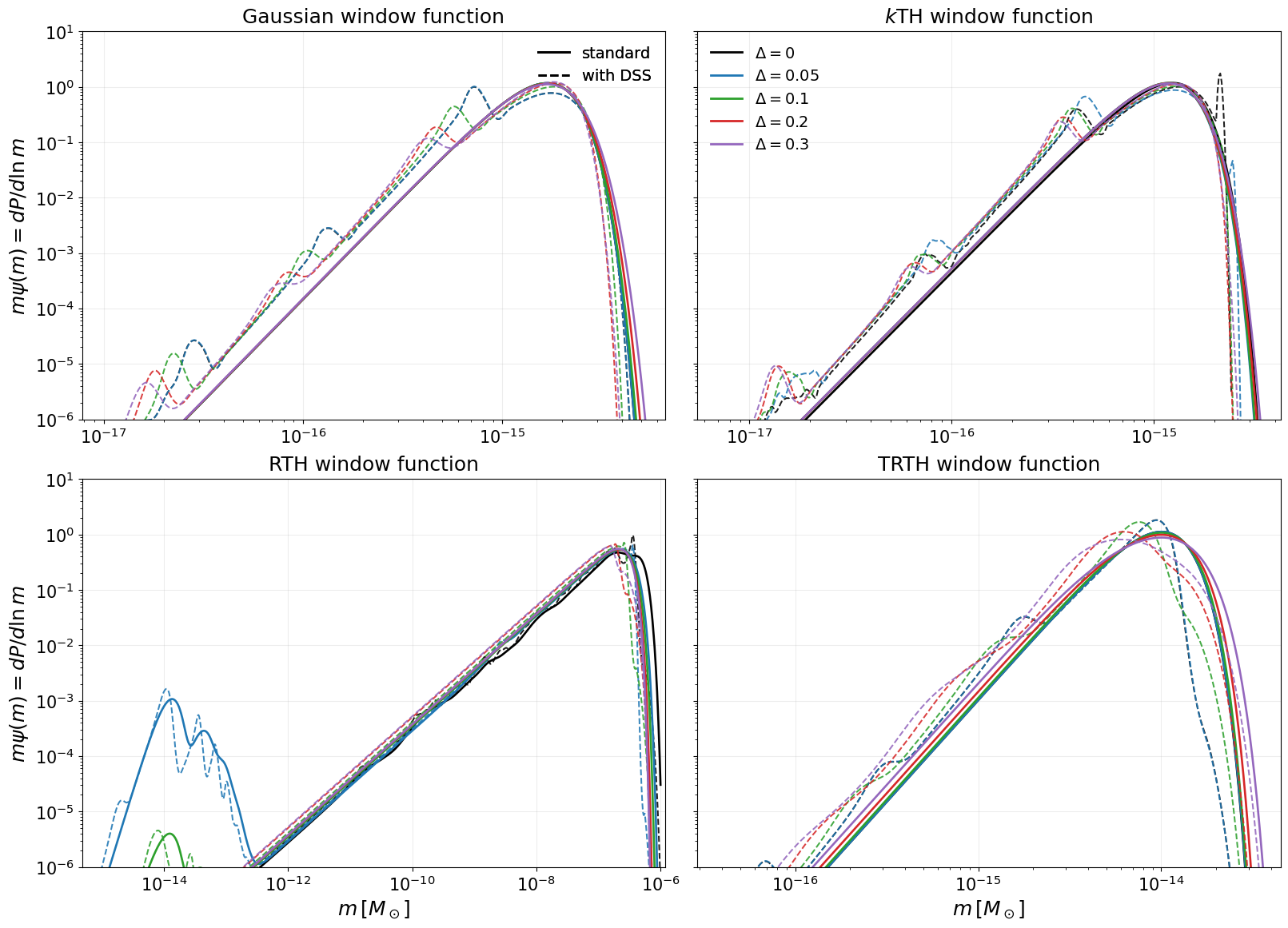}
\caption{\footnotesize{
Comparison of PBH mass functions with and without DSS for narrow lognormal primordial spectra in a kination background. The four panels show $m\psi(m)=\dd P/\dd\ln m$ for Gaussian, $k$TH, RTH and TRTH window functions. Solid curves correspond to {standard power-law} critical scaling and dashed curves include DSS. The primordial widths are $\Delta=0,0.05,0.1,0.2,0.3$. For all finite-width spectra, the maximum height of the primordial curvature spectrum is fixed at $\Pzeta^{\rm peak}=0.05$. The Dirac-delta function reference case is treated separately, with its integrated amplitude chosen to match that of the $\Delta=0.05$ spectrum, $A_\delta=\sqrt{2\pi}(0.05)\Pzeta^{\rm peak}\simeq6.27\times10^{-3}$. For each finite width and {window function}, only the primordial peak position $k_p(\Delta;W)$ is shifted so that the peak of the standard $m\psi_{\rm std}(m)$ distribution coincides with that of the Dirac-delta function reference case. This horizontal alignment removes the overall shift in the characteristic PBH mass and facilitates comparison of the profile width and DSS substructure. After alignment of the standard peaks, the individual DSS features for different values of $\Delta$ need not coincide exactly because finite-width spectra mix contributions from different smoothing scales and horizon masses. Their characteristic logarithmic spacing nevertheless remains controlled by the same DSS period. In the examples shown, the most visible DSS-induced substructures occur mainly on the low-mass side of the distributions. The RTH panel should be read with some caution: Appendix~\ref{app:tophat_domain_sensitivity} shows that its detailed profile is more sensitive to the integration domain than the other window functions.
}}
\label{fig:paper_style_comparison}
\end{figure*}

Figure~\ref{fig:paper_style_comparison} shows the first main comparison. The four panels display the normalized distributions $m\psi(m)$ for Gaussian, $k$TH, RTH and TRTH window functions. Solid curves denote the standard critical-scaling prediction, and dashed curves include DSS.

The main message of Fig.~\ref{fig:paper_style_comparison} is the near overlap of the curves for $\Delta=0,0.05,0.1,0.2,0.3$. In this peak-aligned narrow-spectrum comparison, the normalized mass function is controlled primarily by the critical mass map rather than by the detailed width of the primordial spectrum. The shift of $k_p(\Delta;W)$ removes the overall displacement of the characteristic PBH mass, while the primordial peak height is kept fixed. The narrowness of the spectra prevents strong dephasing over horizon masses. As a result, the standard curves approach a nearly universal mass-function envelope for each window function. This envelope already has a finite width in the Dirac-delta function reference case, because a continuum of supercritical amplitudes above threshold is mapped into a continuum of PBH masses.

DSS is introduced only through the mass map in Eq.~\eqref{eq:dss_scaling}. As already explained, it does not generate the finite width by itself; rather, it modulates the standard mass-function envelope and produces the oscillatory substructure shown by the dashed curves. The persistence of this substructure across the narrow-width sequence shows that the DSS modulation remains coherent for sufficiently narrow primordial spectra. In the examples shown, the most visible DSS-induced structures occur mainly on the low-mass side of the distributions. Their precise location and visibility in the full mass function depend on the statistical weighting and on the integration over smoothing scales and horizon masses.

The individual DSS features are not expected to coincide exactly for different primordial widths. Although the peak-alignment procedure shifts $k_p(\Delta;W)$ so that the maxima of the standard mass functions coincide, a change in $k_p$ by itself would mainly produce a common horizontal translation of both the standard mass-function envelope and the DSS modulation. The remaining relative displacement of the DSS maxima and minima is instead primarily caused by the finite width of the primordial spectrum. For $\Delta>0$, the integration over smoothing scales combines contributions from different horizon masses, each of which maps the same near-threshold {DSS modulation} to a slightly different PBH mass. Their superposition can therefore shift, deform, and partially dephase the visible DSS features, even after the smooth standard peaks have been aligned. This should not be interpreted as a change in the fundamental DSS period.

Figure~\ref{fig:paper_style_comparison} also shows that the DSS imprint is not tied to a particular window function. The Gaussian, $k$TH, and TRTH window functions give qualitatively similar normalized profiles: in all cases the standard mass function provides the smooth finite-width envelope, while the DSS-modulated curve adds log-periodic substructure on top of it. The remaining window function dependence is mostly visible in fine details such as tails, shoulders, the apparent phase, and the relative prominence of individual DSS features, rather than in the existence of the DSS pattern itself.

Among these three more stable {window function} calculations, the DSS substructure is
clearest for the Gaussian and $k$TH window function cases. In the
TRTH window function case, the DSS features remain
present but are less visually pronounced. Although the stiff transfer
function suppresses the slowly decaying large-$kR$ tail of the RTH window function, the combined factor
$W_{\rm TH}(kR)T_{\rm stiff}(kR)$ retains oscillatory power-law support
and is therefore less sharply localized in scale than the Gaussian or $k$TH window functions. The spectral moments can consequently receive
appreciable contributions from a somewhat broader range of modes and
smoothing scales. This produces additional averaging, or partial
dephasing, of the DSS modulation and reduces the contrast of the
individual oscillatory features. The weaker visibility of the DSS
substructure in this panel should therefore be interpreted as a consequence
of the broader scale mixing induced by the window functions,
rather than as an absence of DSS in the underlying critical mass map.

Among the four window function examples, the RTH window function requires the most caution. Although it still exhibits a finite standard mass-function envelope and a DSS-modulated counterpart, its detailed profile is more sensitive to the choice of window function and numerical domain than those obtained with the other windows. We therefore retain it as a useful test of the robustness of the DSS-induced features with respect to the choice of window function, but do not assign physical significance to its detailed peak position, shoulders, or tail structure. Its integration-domain sensitivity is examined explicitly in Appendix~\ref{app:tophat_domain_sensitivity}.

Thus, in this peak-aligned narrow-spectrum comparison, the DSS imprint is more robust than the detailed window-dependent distortions. The window function affects the DSS-induced substructure, phase, and visibility of the modulation, but not the qualitative conclusion that DSS produces log-periodic substructure on top of the minimum width induced by critical scaling, while the location and visibility of individual features remain dependent on the full mass-function convolution.

% ============================================================
\subsection{Broad spectra and loss of critical substructure}
\label{sec:broad_results}

\begin{figure*}[htbp]
\centering
\maybeincludegraphics[width=0.98\textwidth]{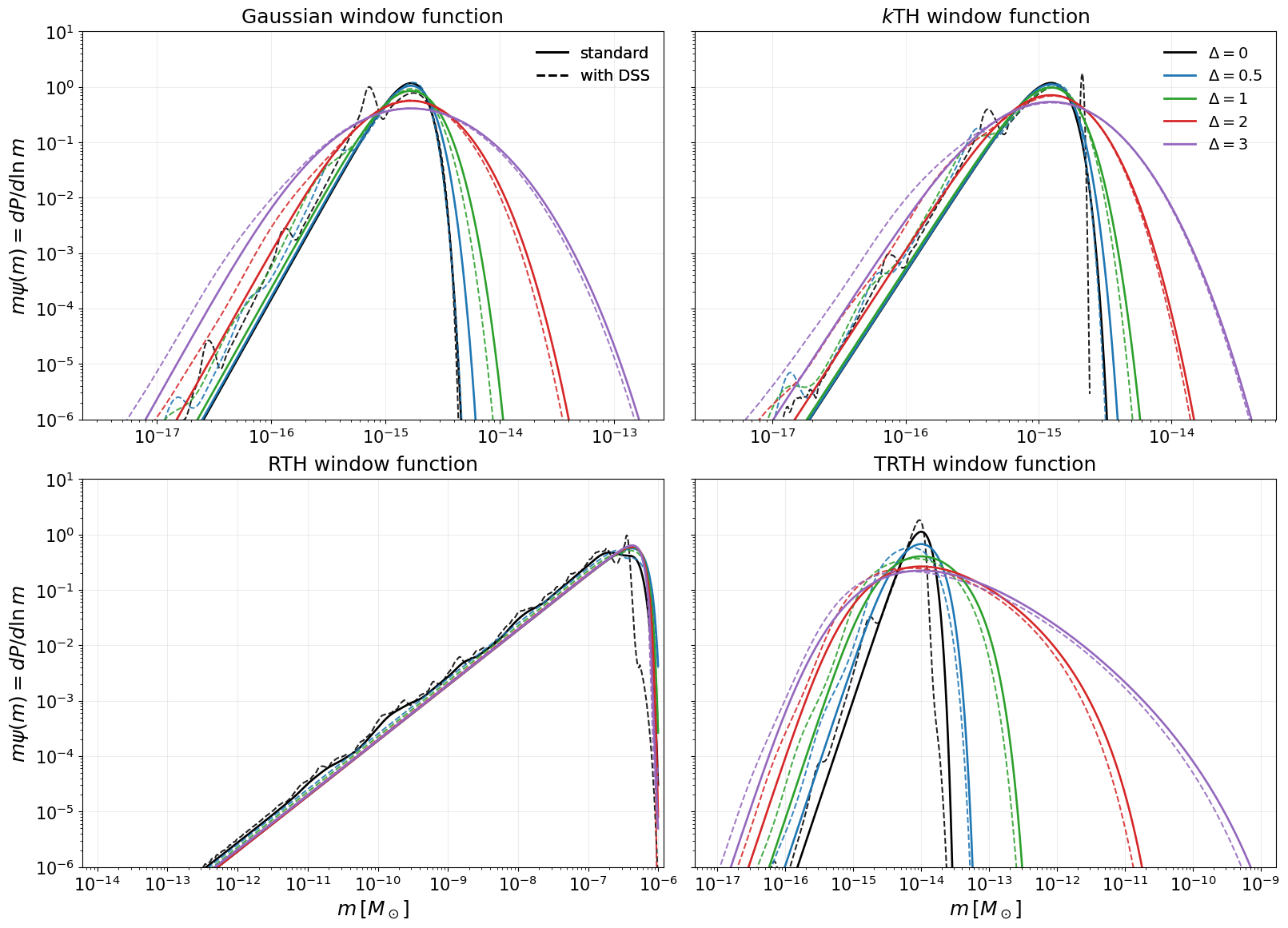}
\caption{\footnotesize{
Width scan for the DSS-modulated PBH mass function in a kination background. The four panels show the normalized distributions $m\psi(m)=\dd P/\dd\ln m$ for Gaussian, 
$k$TH, RTH and TRTH window functions. Solid curves correspond to {standard power-law} critical scaling and dashed curves include DSS. The primordial widths are $\Delta=0,0.5,1,2,3$. For all finite-width spectra, the maximum height of the primordial curvature spectrum is fixed at $\Pzeta^{\rm peak}=0.05$. The Dirac-delta function reference case is treated separately, with its integrated amplitude chosen to match that of the $\Delta=0.05$ spectrum. For each finite width and window function, only $k_p(\Delta;W)$ is shifted so that the peak of the standard $m\psi_{\rm std}(m)$ distribution coincides with that of the Dirac-delta function reference case. As $\Delta$ grows, the integration over smoothing scales mixes contributions from a wider range of horizon masses and progressively suppresses the DSS substructure. The RTH result is strongly sensitive to the upper $R$ boundary and should be interpreted with caution, while the TRTH result is better behaved but still displays a window-induced high-mass distortion.}}
\label{fig:width_scan}
\end{figure*}

One of the most important questions is up to what primordial-spectrum width the DSS imprint remains observable in the PBH mass function. Figure~\ref{fig:width_scan} repeats the same fixed-peak-height prescription used in the narrow comparison but considering the new power spectrum widths $\Delta=0,0.5,1,2,3$. The comparison therefore differs from Fig.~\ref{fig:paper_style_comparison} only in the range of widths explored.

As $\Delta$ grows, the primordial spectrum covers a wider range of scales while maintaining the same peak height. Therefore, the progressive suppression of the DSS substructure is not caused by lowering the maximum power in the primordial spectrum, but by the convolution over a wider range of smoothing scales and horizon masses.

The physical interpretation is phase mixing. For a narrow spectrum, most PBHs form from a relatively narrow range of horizon masses, so the DSS pattern remains coherent in $\ln m$. For a broad spectrum, the same final PBH mass receives contributions from different smoothing scales and therefore from different $M_{\rm H}(R)$. Since the DSS modulation is tied to the near-threshold variable $x$, the integration in Eq.~\eqref{eq:mass_function_integral} superposes contributions in which the same DSS substructure is mapped to different absolute PBH masses. The visible log-periodic features are consequently averaged and progressively washed out.

This washout is clearly visible for the Gaussian, $k$TH, RTH and TRTH window functions. As $\Delta$ increases, the dashed DSS curves approach their corresponding standard mass-function envelopes. The low-mass DSS features that are prominent for the Dirac-delta function reference and moderate widths become much less visible for $\Delta=2$ and $3$. The suppression therefore reflects the broader distribution of horizon masses but not a change in the fundamental DSS period.

For the Gaussian and $k$TH window function cases, the broad mass functions also become increasingly smooth and approach an approximately lognormal-like profile as $\Delta$ grows. In this regime, the shape of the mass function is controlled mainly by the broad lognormal form of the primordial spectrum, while the critical-scaling and DSS substructure become subdominant.

The RTH window function case behaves qualitatively differently. As
discussed in Sec.~\ref{subsec:smoothing_prescriptions} and
Appendix~\ref{app:tophat_domain_sensitivity}, its slowly damped
oscillatory Fourier-space tail makes the spectral moments unusually
sensitive to the smoothing-scale domain. The curves for different
primordial widths consequently develop a similar rise toward large
masses, indicating that their shape is dominated by the window response
rather than by the detailed width of the primordial spectrum.
Contributions from larger smoothing scales are mapped to larger horizon
masses through Eq.~\eqref{eq:horizon_mass_R}, while the final turnover
is controlled by the upper boundary of the $R$ integration. Since this
feature moves when the integration domain is enlarged, its detailed
shape, peak position, and high-mass cutoff should be regarded as domain-dependent rather than as converged physical predictions.

The TRTH window function case is substantially more stable than the RTH window function case, and its DSS substructure is likewise washed out as the primordial spectrum broadens. Nevertheless, the underlying smooth profiles remain visibly distorted and extended toward larger masses relative to the approximately lognormal-like Gaussian and $k$TH window function results. Since this deformation is already present in the standard curves, it is not caused by DSS. It is instead likely associated with the broader oscillatory support of the combined window $W_{\rm TH}(kR)T_{\rm stiff}(kR)$, which gives additional weight to larger smoothing scales and therefore to larger horizon masses through Eq.~\eqref{eq:horizon_mass_R}. The transfer function removes the strong cutoff sensitivity of the RTH window function, but it does not completely eliminate this window-induced high-mass distortion.

This behavior provides a useful way to distinguish two regimes. In the critical-narrow-spectrum limit, the profile is controlled mainly by the near-threshold mass map, and DSS remains visible. In the broad spectrum limit, the width of $\Pzeta(k)$ controls the profile and the integration over horizon masses dephases the DSS modulation. Within the present peak-based construction, the broad-width scan illustrates this transition while also separating the stable window function results from the domain-sensitive RTH window function result.

% ============================================================
\subsection{Quantifying the DSS signal}
\label{sec:dss_diagnostics}

\begin{figure*}[htbp]
\centering
\maybeincludegraphics[width=0.98\textwidth]
{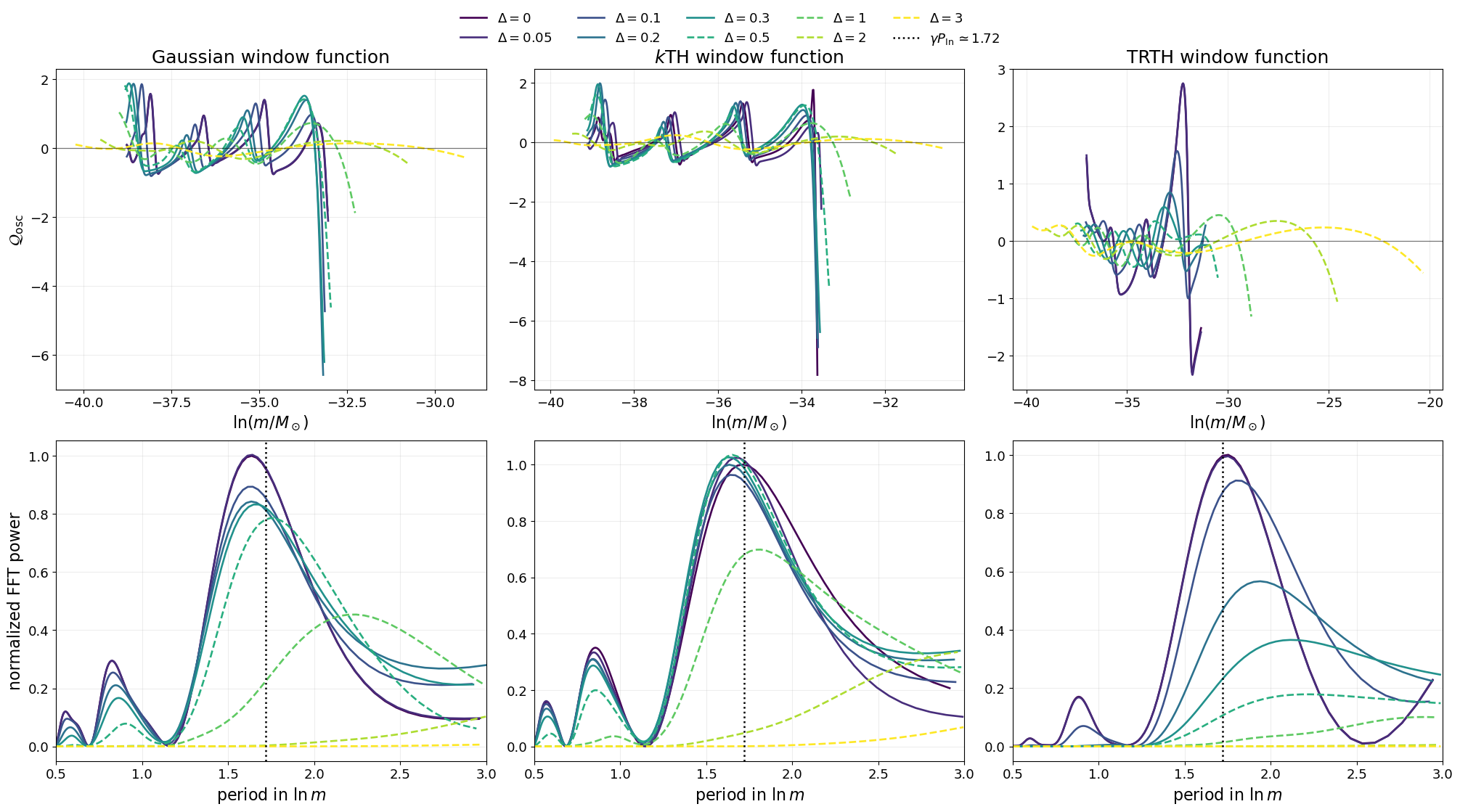}
\caption{\footnotesize{
Residual and periodogram diagnostics for the Gaussian, $k$TH, and TRTH window functions. The upper row shows the detrended logarithmic residual
$\mathcal Q_{\rm osc}=\mathcal Q-\mathcal Q_{\rm smooth}$, where
$\mathcal Q=\ln[(m\psi)_{\rm DSS}/(m\psi)_{\rm std}]$, for all primordial
widths included in the narrow- and broad-spectrum scans. Solid curves correspond
to the narrow sequence
$\Delta=0,0.05,0.1,0.2,0.3$, while dashed curves correspond to the additional
broad-spectrum cases $\Delta=0.5,1,2,3$. The lower row shows the corresponding
FFT periodograms in $\ln m$. Within each column, the periodograms are normalized
to the maximum power of the corresponding $\Delta=0$ reference case. The
vertical dotted lines mark the fixed-horizon-mass expectation
$\Delta\ln m\simeq\gamma P_{\ln}\simeq1.72$. For the Gaussian and $k$TH window function cases, the dominant
periodogram power remains concentrated near the expected DSS scale throughout
the narrow-spectrum regime and persists for moderately broad primordial
spectra. In the TRTH window function case, coherent power
near the same logarithmic scale is also present for narrow spectra, but it is
suppressed and broadened more rapidly as $\Delta$ increases. This faster loss
of coherence is consistent with the broader mixing over smoothing scales and
horizon masses produced by the less localized combined window function
$W_{\rm TH}(kR)T_{\rm stiff}(kR)$. The periodograms are used as coherence
diagnostics rather than as precision measurements of the fundamental DSS
period.
}}
\label{fig:residuals}
\end{figure*}

The previous figures show the DSS signal directly in the normalized mass
functions. To quantify how this signal changes as the primordial spectrum is
broadened, we construct a residual diagnostic from the same numerical output.
For each primordial width and window function, we compute two normalized
mass functions using the same fixed-height amplitude $A(\Delta)$ and the same
aligned value of $k_p(\Delta;W)$: the {standard power-law} critical-scaling result,
$\psi_{\rm std}$, and the DSS-modulated result, $\psi_{\rm DSS}$. We then
define the logarithmic residual
\begin{equation}
\mathcal{Q}(m)
=
\ln\left[
\frac{m\psi_{\rm DSS}(m)}
     {m\psi_{\rm std}(m)}
\right]
=
\ln\left[
\frac{\psi_{\rm DSS}(m)}
     {\psi_{\rm std}(m)}
\right].
\label{eq:dss_log_residual}
\end{equation}
Since both mass functions are normalized, this residual isolates the change in
profile shape induced by the DSS-modulated mass map. In practice, the residual
is evaluated only over the region where both curves remain above a fixed
fraction of their respective maxima. In Fig.~\ref{fig:residuals} we use a tail
cut of $10^{-7}$. This removes the far tails, where the mass functions are
exponentially small and the logarithmic ratio becomes numerically unstable.

The residual $\mathcal Q$ contains both the oscillatory DSS contribution and a
slowly varying component associated with the difference between the broad
profiles of the two mass functions. We therefore subtract a smooth polynomial
trend and define
\begin{equation}
\mathcal{Q}_{\rm osc}(\ln m)
=
\mathcal{Q}(\ln m)
-
\mathcal{Q}_{\rm smooth}(\ln m).
\label{eq:dss_detrended_residual}
\end{equation}
This detrending is used only for the diagnostic in
Fig.~\ref{fig:residuals}; the mass functions shown in
Figs.~\ref{fig:paper_style_comparison} and \ref{fig:width_scan} are not
modified. The detrended residual is then interpolated onto a uniform grid in
$\ln m$, allowing a discrete Fourier transform to be applied.

Figure~\ref{fig:residuals} compares the three window function results that are
most stable under changes of the numerical domain: the Gaussian, $k$TH, and TRTH window function cases. The upper row shows $\mathcal Q_{\rm osc}$ for all
widths used in the narrow- and broad-spectrum scans. For the Gaussian and $k$TH window function cases, the narrow spectra retain a clear and approximately
coherent oscillatory pattern. The residuals for
$\Delta=0,0.05,0.1,0.2,0.3$, and to a lesser extent $\Delta=0.5$, exhibit
similar characteristic separations in $\ln m$, although their detailed phase
and amplitude vary. As the primordial spectrum becomes broader, the residuals
become smoother and their oscillatory contrast decreases. By $\Delta=2$ and
$3$, only a broad, slowly varying deformation remains over most of the reliable
mass range.

The TRTH window function case displays the same
qualitative behavior, but the loss of coherence occurs more rapidly. The
residuals for the narrowest spectra still contain recognizable oscillatory
structure, while moderate widths already show a substantial reduction in
contrast. This is consistent with the broader scale support of the combined window function $W_{\rm TH}(kR)T_{\rm stiff}(kR)$. Although the transfer function
regularizes the slowly decaying tail of the RTH window function, the TRTH window function remains less localized in scale than the Gaussian and $k$TH window functions. The mass-function integral therefore mixes contributions
from a broader range of smoothing scales and horizon masses, producing stronger
dephasing of the DSS pattern.

To characterize the logarithmic scales present in the residuals, we compute
the discrete Fourier transform of $\mathcal Q_{\rm osc}$ on the uniform
$\ln m$ grid using an fast Fourier transform (FFT) algorithm and construct the corresponding
periodogram. The
lower row of Fig.~\ref{fig:residuals} shows the resulting periodograms. For each
window function case, the periodograms are normalized to the maximum power of
the corresponding $\Delta=0$ case, so that the relative suppression with
increasing primordial width can be compared within each column. The vertical
dotted line marks the fixed-horizon-mass DSS expectation
\begin{equation}
P_{\ln m}^{\rm DSS}
\simeq
\gamma P_{\ln}
\simeq 1.72,
\label{eq:expected_dss_period}
\end{equation}
corresponding to the multiplicative mass spacing
\begin{equation}
\frac{m_{n+1}}{m_n}
\simeq
\exp(\gamma P_{\ln})
\simeq 5.6.
\label{eq:expected_dss_mass_ratio}
\end{equation}

For the Gaussian and $k$TH window function cases, the narrow-spectrum
periodograms show a broad concentration of power close to the expected DSS
scale. The peak heights are comparable across the narrow sequence and need not
decrease monotonically at very small $\Delta$, since the periodogram amplitude
depends not only on the coherence of the DSS pattern, but also on the residual
shape, the available logarithmic mass interval, and the redistribution of power
between the fundamental mode and its harmonics. The important result is that
the dominant logarithmic scale remains approximately stable throughout the
narrow-spectrum regime. A clear suppression and broadening of the periodogram
appears only once the primordial spectrum becomes sufficiently wide.

The TRTH window function result shows a broader and more
rapidly suppressed response. The $\Delta=0$ and narrow finite-width cases still
contain substantial power near the expected logarithmic scale, but the peak
shifts and broadens more noticeably as $\Delta$ increases. For moderate and
broad spectra, the periodogram becomes flatter and lower in amplitude than in
the Gaussian and $k$TH cases. This faster loss of coherent power is
consistent with the stronger mixing over smoothing scales inferred from the
mass-function comparison.

The periodograms should not be interpreted as precision measurements of the
fundamental DSS period. The reliable mass interval contains only a small number
of visible oscillations, and the residual is not a pure sinusoid in $\ln m$.
Moreover, the nonlinear critical mass map, the convolution over smoothing
scales, the finite mass interval, and the detrending procedure distribute the
power over a finite range of periods and can shift the location of the discrete
maximum relative to the ideal value $\gamma P_{\ln}$. The periodograms are therefore used
primarily as a coherence diagnostic. With this interpretation,
Fig.~\ref{fig:residuals} confirms that the DSS imprint survives throughout the
narrow-spectrum regime, remains coherent to larger primordial widths for the
Gaussian and $k$TH window functions, and is washed out more rapidly for
the TRTH window function.

% ============================================================
\section{Observational implications}
\label{sec:observational_implications}
The most direct observable consequence of DSS is not a modulation of the primordial curvature spectrum itself, but a modulation in the mapping from primordial fluctuations to PBH masses. Therefore, DSS primarily affects observables that depend on the PBH mass distribution. The distinctive point is that the relevant logarithmic scale is not chosen by the primordial spectrum, but by the critical solution through $\gamma\Pln$. This makes the factor $m_{n+1}/m_n\simeq5.6$ between successive substructures in the mass function a possible phenomenological ruler, provided the primordial spectrum is narrow enough for the DSS phases to remain coherent.

This is the observational continuation of the lesson emphasized by smooth critical behavior mass-function studies: if lognormal templates can be inadequate for narrow peaks because critical behaviour fixes a nontrivial minimum-width shape \cite{Gow:2020cou}, then a DSS critical solution can add still more substructure on top of that smooth mass-function envelope. At the same time, the quantitative size of this effect must be interpreted with the same caution as other window-dependent PBH abundance calculations for broad spectra \cite{Yoo:2020dkz}.

First, abundance constraints on extended PBH mass functions can be locally affected once the overall abundance is fixed by a complete abundance calculation. Even if the integrated PBH fraction is held fixed, DSS can redistribute probability across mass, producing local enhancements and suppressions relative to the {standard power-law} critical-scaling prediction. This may matter in mass regions where constraints vary rapidly with mass, such as evaporation bounds, femtolensing or microlensing limits, and asteroid-mass dark-matter windows.

Second, PBH population observables may inherit the {DSS-induced substructure}. Merger rates, capture rates, and any late-time stochastic background sourced by PBH binaries depend on the PBH mass function and can therefore be affected by log-periodic features. This is distinct from the scalar-induced gravitational-wave background sourced directly by $\Pzeta(k)$: DSS modifies the PBH mass map, not necessarily the primordial spectrum.

Third, the approximate spacing in Eq.~\eqref{eq:dss_spacing_number} provides a distinctive signature. A sequence of features separated by a nearly constant factor in mass would be difficult to mimic with a smooth primordial spectrum alone. However, detecting such a pattern would require the DSS modulation to survive the convolution over horizon masses and observational selection effects.

% ============================================================
\section{Conclusions}
\label{sec:conclusions}
We have studied how discrete self-similarity in the critical behavior of PBH formation affects the PBH mass function. The analysis uses the DSS template motivated by the cosmological scalar-field simulations of Ref.~\cite{Padilla:2026dss}, where the residuals around the critical mass-scaling law were found to be log-periodic. The key point is that the {standard power-law} critical-scaling relation already imposes a minimum width on the mass function, even for a Dirac-delta function primordial spectrum. DSS then modulates this irreducible profile and produces approximately log-periodic substructure in mass.

At {a} fixed horizon mass, successive equal-phase points of the DSS-modulated critical mass map obey
\begin{equation}
  \Delta\ln M_{\rm PBH}=\gamma\Pln.
\end{equation}
After constructing the full mass function, this provides the reference expectation
\begin{equation}
  \Delta\ln m\simeq\gamma\Pln,
\end{equation}
which gives $m_{n+1}/m_n\simeq5.6$ for the fiducial Choptuik DSS parameters used here. The first relation is exact for the fixed-horizon-mass critical map, whereas the second describes the approximate spacing of visible substructures after the convolution over perturbation amplitudes and smoothing scales.

The visibility of the signal depends strongly on the width of the primordial spectrum. For narrow spectra, the DSS pattern remains coherent and can produce pronounced substructure. For broad spectra with fixed peak height, the integration over horizon masses mixes different phases of the DSS oscillation and progressively washes out the critical-behavior-induced features. Within the present peak-based construction, this shows that the loss of DSS substructure is primarily a dephasing effect rather than a consequence of lowering the maximum amplitude of the primordial spectrum as the width is increased. Window functions affect the detailed amplitudes, phases, and tails, but the qualitative distinction between the narrow-spectrum limit and broad-spectrum regimes persists across the stable {window functions} considered here. The RTH window function is the least stable {window function} in this comparison: without transfer function suppression of its oscillatory high-$kR$ tail, the detailed mass-function profile can depend appreciably on the adopted integration domain. For this reason, we use it primarily as a sensitivity test and give greater weight to conclusions that are common to the Gaussian, $k$TH, and TRTH window functions.

Because the present calculation uses normalized mass functions, the main
conclusions concern the profile shape rather than the absolute abundance. The
fiducial effective density threshold is inferred from the maximum-compaction
threshold reported in Ref.~\cite{Padilla:2026dss}, using the long-wavelength
relation between the maximum compaction and the extrapolated volume-averaged
density contrast. Its precise 
value in terms of the smoothed-density
variable entering the peak statistics is relevant for absolute-abundance
predictions.

Future work should connect this simplified treatment to mass functions extracted directly from numerical collapse simulations, include non-Gaussian primordial statistics, and propagate DSS-modulated mass functions into observational constraints and PBH population predictions. It will also be important to investigate how nonsphericity, anisotropy, and angular momentum affect the scalar-field PBH threshold and abundance in a kination background. These effects are important for a fully generic treatment of PBH formation, but they would not alter the main point studied here: whenever the near-threshold scalar-field mass map is governed by a DSS critical solution, the associated {DSS modulation} should be propagated into the normalized PBH mass-function profile.

% ============================================================

\appendix

\section{Integration-domain sensitivity of the window functions}
\label{app:tophat_domain_sensitivity}

The mass-function construction requires finite numerical domains for the
integrations over wavenumber, smoothing scale, and distance to threshold.
The main-text results use the fiducial smoothing-scale interval defined in
Eq.~\eqref{eq:fiducial_R_domain}. In this appendix, we examine explicitly
whether the normalized mass-function profiles are stable when this interval
is enlarged. The purpose is not to introduce a different physical setup, but
to distinguish features that have converged with respect to the numerical
domain from features that track the adopted smoothing-scale boundaries.

This test is particularly important for the RTH window function.
From Eq.~\eqref{eq:tophat_window}, its large-$z$ behavior is
\begin{equation}
W_{\rm TH}(z)
\simeq
-\frac{3\cos z}{z^2},
\qquad z\gg1.
\label{eq:appendix_tophat_asymptotic}
\end{equation}
Consequently, the combination entering the density power spectrum satisfies
\begin{equation}
z^4W_{\rm TH}^2(z)
\simeq
9\cos^2 z.
\label{eq:tophat_large_z_kernel}
\end{equation}
The factor $z^4$ in Eq.~\eqref{eq:pdelta_smoothed} therefore compensates the
leading power-law decay of the window function. The resulting effective
weight remains oscillatory but is not asymptotically suppressed. After the
wavenumber integration, this can leave appreciable support over a wide range
of smoothing scales and make the spectral moments, and hence the normalized
mass-function profile, sensitive to the upper boundary of the $R$ integral.

By contrast, the kination transfer function behaves asymptotically as
$T_{\rm stiff}(z)\propto z^{-3/2}$ up to oscillations. The
TRTH window function therefore satisfies
\begin{equation}
z^4W_{\rm TH}^2(z)T_{\rm stiff}^2(z)
\propto z^{-3},
\label{eq:tophat_transfer_large_z_kernel}
\end{equation}
which supplies additional large-$z$ suppression. {It is, however, noted 
that the second 
and higher spectral moments are still sensitive to the upper boundary of the $R$ integral based on Eq.~\eqref{eq:sigma_moments}
even with the TRTH window function.}
The Gaussian and $k$TH window functions are, on the other hand, more localized in scale and are
therefore expected to be less sensitive to an enlargement of the
smoothing-scale domain.

To test this expectation, we repeat the mass-function calculation for a
representative narrow primordial spectrum with $\Delta=0.1$. We use the same
primordial amplitude and the same aligned value of $k_p(\Delta;W)$ as in
Fig.~\ref{fig:paper_style_comparison}. Importantly, $k_p(\Delta;W)$ is not retuned
when the integration domain is changed, since such a retuning would
remove part of the domain-induced displacement that the test is intended to
measure. All other numerical and physical parameters are held fixed.

We consider the three smoothing-scale intervals
\begin{subequations}
\label{eq:appendix_R_domains}
\begin{align}
\mathcal D_1:\quad&
\frac{3\times10^{-3}}{k_p(\Delta;W)}
\leq R\leq
\min\left[
\frac{3\times10^3}{k_p(\Delta;W)},
R_{\rm kin}
\right],
\\
\mathcal D_2:\quad&
\frac{3\times10^{-4}}{k_p(\Delta;W)}
\leq R\leq
\min\left[
\frac{3\times10^4}{k_p(\Delta;W)},
R_{\rm kin}
\right],
\\
\mathcal D_3:\quad&
\frac{3\times10^{-5}}{k_p(\Delta;W)}
\leq R\leq
\min\left[
\frac{3\times10^5}{k_p(\Delta;W)},
R_{\rm kin}
\right].
\end{align}
\end{subequations}
The last interval is the fiducial domain used for the main-text figures. The
number of smoothing-scale points is increased with the logarithmic size of
the domain so that the sampling density in $\ln R$ remains approximately
fixed. The wavenumber and near-threshold integrations are chosen sufficiently
wide to contain the support required by all three cases.

\begin{figure*}[htbp]
\centering
\maybeincludegraphics[width=0.98\textwidth]
{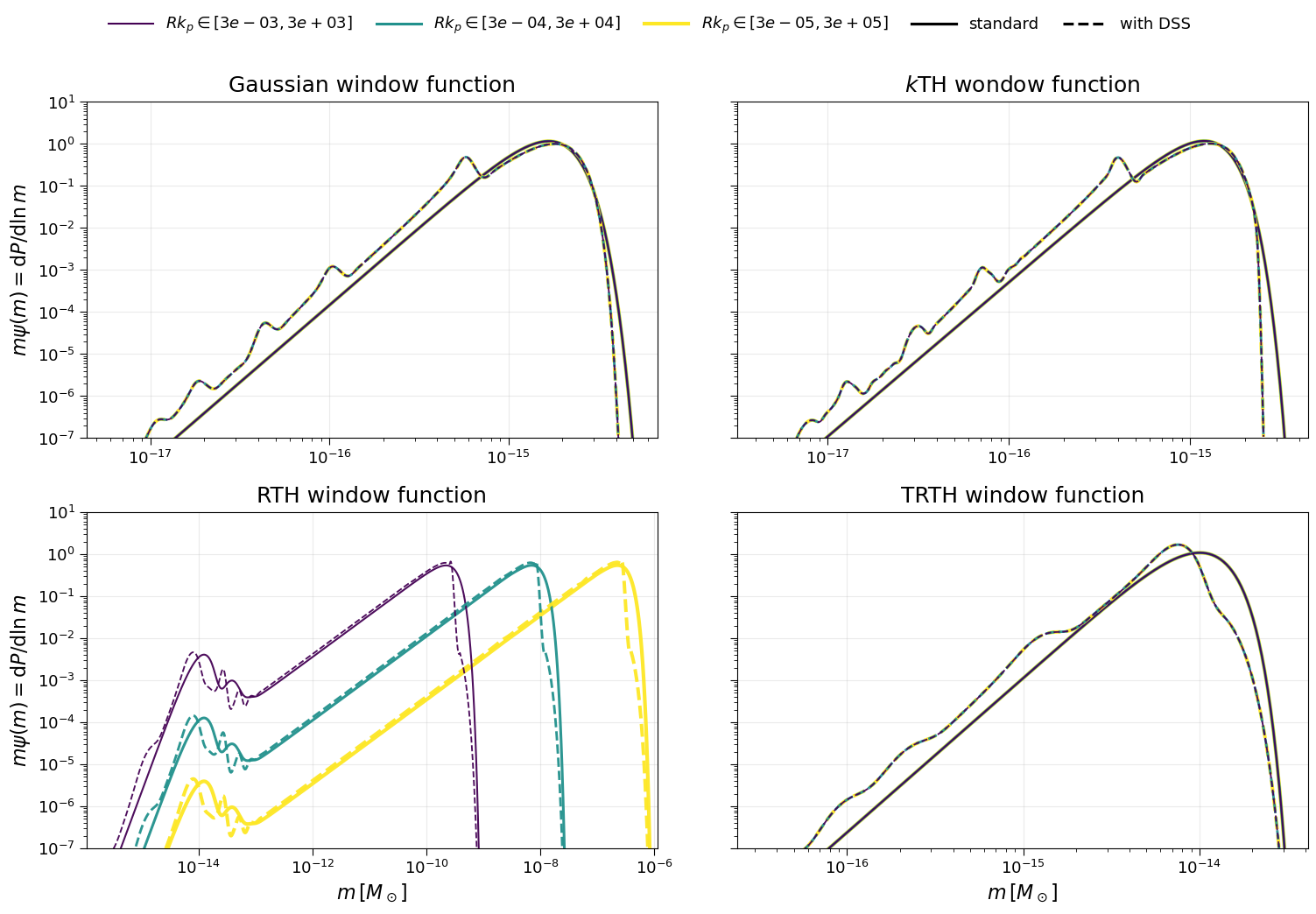}
\caption{\footnotesize{
Integration-domain test for the four window function examples, using the
representative primordial width $\Delta=0.1$. The panels show the normalized
mass functions $m\psi(m)$ for the Gaussian, $k$TH, RTH, and TRTH window functions. Different colors correspond to the three nested smoothing-scale
domains $\mathcal D_1$, $\mathcal D_2$, and $\mathcal D_3$ defined in
Eq.~\eqref{eq:appendix_R_domains}. Solid curves show {standard power-law} critical mass
scaling, while dashed curves include the DSS modulation. The primordial
amplitude, the aligned value of $k_p(\Delta;W)$, and all remaining physical parameters
are kept fixed as the domain is enlarged. The Gaussian, $k$TH, and TRTH results are nearly unchanged. In contrast, the RTH window function profile case changes substantially: its peak and
high-mass turnover move as the upper $R$ boundary is increased. This
demonstrates that the detailed high-mass structure of the RTH window function result is controlled by the integration domain rather than by the DSS
mass map.
}}
\label{fig:domain_sensitivity_all_windows}
\end{figure*}

Figure~\ref{fig:domain_sensitivity_all_windows} compares the standard and
DSS-modulated mass functions obtained with these domains. For the Gaussian, 
$k$TH, and TRTH window function cases, the curves from the three domains nearly
overlap over the mass range carrying appreciable probability. Their peak
positions, widths, and DSS substructure are therefore stable under the domain
enlargement.

The RTH window function case behaves differently. Enlarging the upper
smoothing-scale boundary changes the relative weight assigned to large $R$,
which corresponds to larger horizon masses through
Eq.~\eqref{eq:horizon_mass_R}. Its peak position, high-mass rise, and final
turnover consequently move as the domain is enlarged. The turnover follows
the upper boundary of the $R$ integration rather than approaching a fixed
mass, demonstrating that this part of the profile is cutoff dependent. Since
the same behavior is present in both the standard and DSS-modulated curves,
it is a property of the {RTH window function and integration domain} rather
than an effect generated by DSS.

This comparison supports the interpretation adopted in the main text. The
minimum width generated by critical mass scaling and the existence of
DSS-induced substructure are stable for the Gaussian, $k$TH and TRTH window functions.
By contrast, the detailed peak position and high-mass structure of the RTH window function result should not be regarded as converged physical
predictions within the finite-domain calculation.
% ============================================================
\begin{acknowledgments}
L.E.P. acknowledges support from JSPS KAKENHI Grant Number JP25KF0278. T.H. acknowledges support from JSPS KAKENHI Grant Number JP24K07027. H.I acknowledges support from Rikkyo University Special Fund for Research.
% This work waspartially supported by Rikkyo University Special Fund for Research (H.I.)
\end{acknowledgments}

% ============================================================
% Bibliography. Replace placeholder entries with exact BibTeX keys if using a .bib file.
% The current version is self-contained and compiles without an external biblio.bib.
% ============================================================

\bibliographystyle{ieeetr}
\bibliography{biblio}

\end{document}